\documentstyle[aps,prl,epsfig,graphics]{revtex}

\begin{document}

\draft

\twocolumn[\hsize\textwidth\columnwidth\hsize\csname
@twocolumnfalse\endcsname

\title{Operationally Invariant Information in Quantum Measurements}
\author{\v Caslav Brukner and Anton Zeilinger}
\address{Institut f\"ur Experimentalphysik, Universit\"at Wien,\\
         Boltzmanngasse 5, A--1090 Wien, Austria}

\maketitle


\vskip1pc]

\begin{abstract}

A new measure of information in quantum mechanics is proposed
which takes into account that for quantum systems the only feature
known before an experiment is performed are the probabilities for
various events to occur. The sum of the individual measures of
information for mutually complementary observations is invariant
under the choice of the particular set of complementary
observations and conserved if there is no information exchange
with an environment. That operational quantum information
invariant results in $k$ bits of information for a system
consisting of $k$ qubits.

\end{abstract}



In any individual quantum measurement with discrete variables a
number of different outcomes are possible, for example, in a
spin-1/2 measurement the individual outcomes ''spin up'' and
''spin down''. We define a new measure of information for an
individual quantum measurement based on the fact that the only
feature defined before the measurement is performed are the
specific probabilities for all possible individual outcomes.

The observer is free to choose different experiments which might
even completely exclude each other, for example, measurements of
orthogonal components of spin. This quantum complementarity of
variables occurs when the corresponding operators do not commute.
One quantity, for example, the $z$ component of spin, might be
well defined at the expense of maximal uncertainty about the other
orthogonal components. We define the total information content in
a quantum system to be the sum over all individual measures for a
complete set of mutually complementary experiments.

The experimentalist may decide to measure a different set of
complementary variables thus gaining certainty about one or more
variables at the expense of loosing certainty about other(s). In
the case of spin these could be the projections along rotated
directions, for example, where the uncertainty in one component is
reduced but the one in another component is increased
correspondingly.

Intuitively one expects that the total uncertainty or,
equivalently, the total information carried by the system is {\it
invariant} under such transformation from one complete set of
complementary variables to another. We show that the total
information defined according to our new measure has exactly that
invariance property. Also it is conserved in time if there is no
information exchange with an environment.

We find that the total information of a system results in $k$ bits
of information for a system consisting of $k$ qubits. For a
composite system, maximal entanglement results if the total
information carried by the system is exhausted in specifying joint
properties, with no individual qubit carrying any information on
its own. Our results we interpret as implying that information is
the most fundamental notion in quantum mechanics.

Every reasonably well-designed experiment tests some {\em
proposition}. Knowledge of the state of a quantum system permits
the prediction of individual outcomes with certainty only for that
limited class of experiments which have definite outcomes, a
situation where the corresponding propositions have definite truth
value. From theorems like Kochen-Specker \cite{kochen} we know
that in quantum mechanics it is not possible, not even in
principle, to assign definite noncontextual truth values to all
conceivable propositions. About indefinite propositions we can
make only probabilistic predictions.

Consider a stationary experimental arrangement with $n$ possible
outcomes. Knowing the probabilities
$\vec{p}=(p_1,...,p_j,...,p_n)$ for the outcomes {\it all} an
experimenter can do is to guess how many times a specific outcome
will occur. In making his prediction he has only a limited number
of systems to work with. Then, because of the statistical
fluctuations associated with any finite number of experimental
trials, the number $n_j$ of occurrences of a specific outcome $j$
in future $N$ repetitions of the experiment is not precisely
predictable. Rather, the experimenter's uncertainty
(mean-square-deviation), or lack of information, in the value
$n_j$ is \cite{gnedenko}
\begin{equation}
\sigma_j^2 = p_j (1-p_j) N. \label{error}
\end{equation}
This implies that for a sufficiently large number $N$ of
experimental trials the confidence interval is given as
$(p_jN\!-\!\sigma_j,p_jN\!+\!\sigma_j)$. Therefore, if we just
plan to perform the experiment $N$ times, we know {\em in
advance}, before the experiments are performed and their outcomes
become known, that the number $n_j$ of future occurrences of the
outcome $j$ will be found with probability 68\% within the
confidence interval.

Notice that the experimenter's lack of information (\ref{error})
is proportional to the number of trials. This important property
guarantees that each individual performance of the experiment
contributes the same amount of information, no matter how many
times the experiment has already been performed. After each trial
the experimenter's lack of information about the outcome $j$
therefore decreases by
\begin{equation}
U(p_j) = \frac{\sigma^2_j}{N}=p_j(1-p_j).
\end{equation}
This is the lack of information about the outcome $j$ with respect
to a single future experimental trial. In this view we suggest to
define the total lack of information regarding all $n$
possible experimental outcomes as
\begin{equation}
U(\vec{p}) =\sum_j^{n} U(p_j) =
\sum_j^{n}p_j(1-p_j)=1-\sum_j^{n} p^2_j.
\end{equation}
The uncertainty is minimal if one probability is equal to one and
it is maximal if all probabilities are equal.

This suggests that the knowledge, or information, with respect to
a single future experimental trial an experimentalist possesses
before the experiment is performed is a complement of $U(\vec{p})$
and, furthermore, that it is a function of a sum of the squares of
probabilities. A first ansatz therefore would be $I(\vec{p}) = 1 -
U(\vec{p}) = \sum_{i=1}^{n} p^2_i$. Expressions of such a general
type were studied in detail by Hardy, Littlewood and P\'{o}lya
\cite{hardy}. Notice that this expression can also be viewed as
describing the length of the probability vector $\vec{p}$.
Obviously, because of $\sum_{i} p_i=1$, not all vectors in
probability space are possible. Indeed, the minimum length of
$\vec{p}$ is given when all $p_i$ are equal ($p_i=1/n$). This
corresponds to the situation of complete lack of information in an
experiment. Therefore we suggest to normalize the measure of
information in an individual quantum experiment as obtaining
finally
\begin{equation}
I(\vec{p})=  {\cal N} \sum_{i=1}^{n} \left( p_i-\frac{1}{n}
\right)^2.
\label{junko}
\end{equation}
Considering from now on those cases where maximally $k$ bits of
information can be encoded, i.e., $n=2^k$, the normalization is
${\cal N} = 2^k k/(2^k-1)$. Then $I(\vec{p})$ results in $k$ bits
of information if one $p_i=1$ and it results in 0 bits of
information when all $p_i$ are equal.

We emphasize that our measure of information is not equal to
Shannon's information. While Shannon's information is applicable
when measurement reveals a preexisting property \cite{caslav2}, in
general, a quantum measurement does not reveal a preexisting
property.

Having defined the measure of information for an individual
quantum measurement we now ask what the total information content
in a quantum system is. We recall Bohr's\cite{bohr} remark that
''... phenomena under different experimental conditions, must be
termed complementary in the sense that each is well defined and
that together they exhaust all definable knowledge about the
object concerned,'' and suggest to {\em sum} the individual
measures of information [Eq. \ref{junko}] over a complete set of
$m$ of mutually complementary experiments
\begin{equation}
I_{total} = \sum_{j=1}^{m} I_j(\vec{p}).
\end{equation}

A set of propositions associated with certain quantum-mechanical
experiments is {\em mutually complementary} if complete knowledge
of the truth value of any one of the propositions implies {\em
maximal} uncertainty about the truth values of the others. Such a
complete set of propositions for a spin-1/2 particle can be for
example the following: (1) ''The spin along the $x$ axis is up'',
(2) ''The spin along the $y$ axis is up'' and (3) ''The spin along
the $z$ axis is up'' \cite{comment}.

\begin{figure}
\centerline{\psfig{width=5.5cm,file=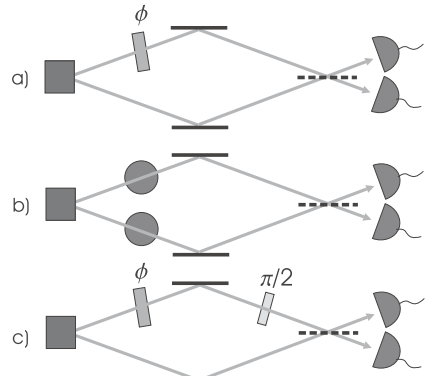}}
\caption{Principle sketch of arrangements to consider mutually
exclusive classes of information in an interference experiment
with a Mach-Zehnder type of interferometer. Into each of the two
paths of the interferometer in (b) one detector is inserted with a
property to detect the particle without absorbing it.}
\label{mach}
\end{figure}

Another example for complementary is quantum interference.
Consider an experiment with an idealized Mach-Zehnder type of
interferometer (Fig. \ref{mach}). Suppose that for a specific
phase shift $\phi$ between the two beams inside the interferometer
(Fig. \ref{mach}a), the particle will exit with certainty towards
the upper (lower) detector behind the beam splitter. In this case
we have complete knowledge of the beam the particle will be found
in behind the beam splitter at the expense of the fact that we
have absolutely no knowledge which path the particle took inside
the interferometer. The state of the particle is then represented
by the truth value (true or false) of the proposition: (1) ''The
particle takes the outgoing path towards the upper detector in
presence of the phase shift $\phi$.''

In contrast, if we know which path the particle took through the
interferometer (Fig. \ref{mach}b) no interference results and
hence it is completely uncertain which outgoing path the particle
will take. The state of the particle can now be specified by the
truth value of the proposition: (2) ''The particle takes the upper
path inside the interferometer.''

Knowing that spin-1/2 affords a model of the quantum mechanics of
all two-state systems, i.e. qubits, we expect that there are
always three mutually complementary propositions whenever binary
alternatives are considered. Indeed, it can easily be shown that
even without path information our knowledge of the beam the
particle will be found in behind the beam splitter in Fig.
\ref{mach}a will be completely removed if we introduce an
additional phase shift of $\pi/2$ between the two beams inside the
interferometer. Then, in the new arrangement in Fig. \ref{mach}c
both outgoing beams will be equally probable.

Now, suppose that in the presence of a specific phase shift
$\phi+\pi/2$ (Fig. \ref{mach}c), the particle will exit with
certainty towards the upper (lower) detector. The state of the
system is now represented by the truth value of the proposition
(3) ''The particle takes the outgoing path towards the upper
detector in presence of the phase shift $\phi\!+\!\pi/2$''. For a
particle in that state we have complete knowledge of the outgoing
beam the particle will take (Fig. \ref{mach}c) at the expense of
absolutely no knowledge about either the path inside the
interferometer (Fig. \ref{mach}b) or about the outgoing path in
the arrangement in Fig. \ref{mach}a.

Notice that we can label various sets of the three mutually
complementary propositions by the value $\phi$ of the phase shift.
The three propositions we found for the interferometer are
formally equivalent to the complementary propositions about
spin-1/2: (1)''The spin is up along $\phi$ in the $x$-$y$ plane'',
(2) ''The spin is up along the $z$ axis'', and (3) ''The spin is
up along $\phi+\pi/2$ in the $x$-$y$ plane''. Here, the direction
$\phi$ is assumed to be by lying in the $x$-$y$ plane oriented at
an angle $\phi$ with respect to the $x$ axis. Evidently, this
analogy can be carried even further using the concept of
multiports. Therefore from now on we will explicitly discuss spin
measurements only keeping in mind the applicability of these ideas
for interference experiments.

We realize that the total information content of the system is
\begin{eqnarray}
I_{total} &=& I_1(p^+_1,p^-_1)+I_2(p^+_2,p^-_2)+I_3(p^+_3,p^-_3)
\nonumber
\\ &=& 2Tr\hat{\rho}^2-1. \label{tic1}
\end{eqnarray}
Here, e.g., $p^+_1$ is the probability to find the particle in the
state $\hat{\rho}$ with spin up along $\phi$. Evidently, this is
{\em invariant} under {\em unitary} transformations. Also, this
results in just 1 bit of information for a pure state when one
single proposition with definite truth value is assigned to the
system and in 0 bits of information for a completely mixed state
when no proposition with definite truth value can be made about
the system.

Note that the total information content of a quantum system is
completely specified by the state of the system alone and {\em
independent} of the physical parameter $\phi$ (phase shift) that
labels various sets of mutually complementary observations. In the
same spirit as choosing a coordinate system one may choose any set
of mutually complementary propositions to represent our knowledge
of the system. The total information about the system will then be
invariant under that choice. This is the reason we may use the
phrase ''the total information content'' without explicitly
specifying the particular reference set of mutually complementary
propositions. Also note that the total information content of the
system is {\em conserved} in time if there is no information
exchange with the environment, that is, if the system is
dynamically independent from the environment and not exposed to a
measurement.

Wootters and Zurek \cite{zurek} found for a double-slit experiment
that we can obtain some partial knowledge about the particle's
path and still observe an interference pattern of reduced contrast
as compared to the ideal interference situation. Englert
\cite{englert} has proposed an inequality to describe
quantitatively the complementarity between path information and
interference pattern in a Mach-Zehnder type of interferometer. Our
results indicate that when we take into account not just two
variables, but three, the rigorous equality Eq. (\ref{tic1})
results.

In order to analyze the most simple composite system in view of
the ideas just proposed above, let us consider two qubits. An
explicit example will again be two spin-1/2 particles. We will
consider a complete set of mutually complementary {\it pairs} of
propositions where precise knowledge of the truth values of a
specific pair of propositions excludes any knowledge of the truth
values of other complementary pairs of propositions. As opposed to
the single-particle case where three individual propositions are
complementary to each other, in the two-particle case we have five
pairs of propositions where each pair is complementary to each
other pair \cite{wootters}.

We give one possible choice of a complete set of five pairs of
complementary propositions for two particles: (1) ''The spin of
particle 1 is up along $z$'' and ''The spin of particle 2 is up
along $z$;'' (2) ''The spin of particle 1 is up along $\phi_1$''
and ''The spin of particle 2 is up along $\phi_2$'', (3) ''The
spin of particle 1 is up along $\phi_1+\pi/2$'' and ''The spin of
particle 2 is up along $\phi_2+\pi/2$,'' (4) ''The spin of
particle 1 along $z$ and the spin of particle 2 along $\phi_2$ are
the same'' and ''The spin of particle 1 along $\phi_1$ and the
spin of particle 2 along $\phi_2+\pi/2$ are the same,'' (5) ''The
spin of particle 1 along $z$ and the spin of particle 2 along
$\phi_2+\pi/2$ are the same'' and ''The spin of particle 1 along
$\phi_1$ and the spin of particle 2 along $z$ are the same.''
Again directions $\phi_1$ and $\phi_2$ are assumed both to be by
lying in the $x$-$y$ plane oriented at an angle $\phi_1$ and
$\phi_2$, respectively, with respect to the $x$ axis. In a set of
mutually exclusive two-particle interference experiments the
angles would correspond to phase shifts in the two-particle
interferometer.

We find for the total information carried by the composite system
\begin{equation}
I_{total} = \sum_{j=1}^{5} I_j(\vec{p}^j) =
\frac{2}{3}(4Tr\hat{\rho}^2-1).
\label{tic2}
\end{equation}
Here, $\vec{p^j}=(p^j_1,p^j_2,p^j_3, p^j_4)$ are the probabilities
for the system in the state $\hat{\rho}$ to give the four possible
combinations (true-true, true-false, false-true and false-false)
of the truth values for the pair $j$ of propositions. This again
is invariant under unitary transformations. Independence of
physical parameters $\phi_1$ and $\phi_2$ implies that the total
information of the composite system is invariant under the choice
of the particular set of five mutually complementary pairs of
propositions. Also the total information of the composite system
is conserved in time if there is no information exchange between
the composite system and an environment. We note that these
results can be generalized to a composite system consisting of $k$
qubits.

A composite 2-qubits system in a pure state carries 2 bits of
information. That information contained in two propositions can be
distributed over the two particles in various ways. It may be
carried by the two particles individually, e.g., as the two-bit
combination false-true of the truth values of the propositions
given in (1). This is then represented by the product state $
|\psi\rangle_{prod}=|z-\rangle_1 |z+\rangle_2$. The 2 bits of
information are thus encoded in the two particles separately, one
bit in each particle just as in classical physics. In that case
there is no additional information represented jointly by the two
systems.

Alternatively, 2 bits of information might all be carried by the
two particles in a joint way, in the extreme with no individual
particle carrying any information on its own. For example, this
could be the two-bit combination true-false of the truth values of
the propositions given in (4). Again, this is represented by the
entangled state
\begin{eqnarray}
|\psi\rangle_{ent} &=& \frac{1}{\sqrt{2}} ( i |z+\rangle
|x(\phi_2)+\rangle + |z-\rangle |x(\phi_2)-\rangle)
\\ &=& \frac{1}{\sqrt{2}} (|x(\phi_1)+\rangle |y(\phi_2)-\rangle - i
|x(\phi_1)-\rangle |y(\phi_2)+\rangle), \nonumber
\end{eqnarray}
where, e.g., $|x(\phi)+\rangle$ represents the eigenstate spin-up
along a direction rotated by $\phi_1$ from $x$. This Bell state
does not contain any information about the individuals; all
information is contained in joint properties. In fact, now there
cannot be any information carried by the individuals because the
two bits of information are exhausted by defining that maximally
entangled state, and no further possibility exists to also encode
information in individuals. This we see as a quantitative
formulation of Schr\"{o}dinger's \cite{schroedinger} idea that '' If
two separated bodies, each by itself known maximally, enter a
situation in which they influence each other, and separate again,
then ... the knowledge remains maximal, but at its end, if the two
bodies have again separated, it is not again split into a logical
sum of knowledges about the individual bodies.''

For clarity we emphasize that our total information content of a
quantum system is neither mathematically nor conceptually
equivalent to von Neumann's entropy. With the only exception for
results of measurement in a basis decomposing the density matrix
into a classical mixture when it can be considered as equivalent
to Shannon's information, the von Neumann entropy is just a
measure of the purity of the given density matrix without explicit
reference to information contained in individual measurements
\cite{caslav2}. In contrast, our information content is purely
operational and refers directly to experimental results of
mutually complementary measurements, thus including also those for
which the density matrix cannot be decomposed into a classical
mixture. Our information content of the system can be viewed as
equivalent to the sum of partial knowledges an experimentalist can
have about mutually exclusive measurements without any further
reference to the structure of the theory.

In the present paper we find an operational quantum information
invariant that reflects the intrinsic symmetry of the underlying
Hilbert space of the system. We interpret our result as implying
that number of essential features of quantum mechanics, might be
based on the observation \cite{zeilinger7,malus} that {\em N most
elementary systems represent the truth values of N propositions
only}. Since this is the only information quantum systems carry, a
measurement associated with any other proposition must necessarily
contain an element of randomness. This kind of randomness must
then be irreducible, that is, it cannot be reduced to ''hidden''
properties of the system. Otherwise the system would carry more
information.

Entanglement results from the fact that information could also be
distributed in joint properties of a multiparticle system. In
particular, maximal entanglement arises when the total information of a
composite system is exhausted in specifying joint
properties.

This work have been supported by Austrian Science Foundation FWF,
Project No. S6503 and No F1506.


\begin{thebibliography}{99}
\bibitem{comment} These statements we only mean to be statements
about detector clicks in Stern-Gerlach type experiments.
\end{thebibliography}
\end{document}